\documentclass[a4paper]{article}

\usepackage[english]{babel}
\usepackage[utf8x]{inputenc}
\usepackage[T1]{fontenc}
\usepackage{afterpage}

\usepackage[a4paper,top=3cm,bottom=2cm,left=3cm,right=3cm,marginparwidth=1.75cm]{geometry}

\usepackage{amsmath}
\usepackage{graphicx}
\usepackage{color}
\usepackage{cancel}
\usepackage[normalem]{ ulem }
\usepackage{soul}
\usepackage[colorinlistoftodos]{todonotes}
\usepackage{pdfpages}
\usepackage{enumerate}
\usepackage{subcaption}
\usepackage{authblk}

\title{Measuring Nestedness: \\ A comparative study of the performance of different metrics}

\author[1,2]{Cl\`audia Payrat\'o\--Borr\`as}
\author[1]{Laura Hern\'andez}
\author[2,3,4]{Yamir Moreno}

\affil[1]{ Laboratoire de Physique Th\'eorique et Mod\'elisation, UMR CNRS, Universit\'e de Cergy-Pontoise, 2 Avenue Adolphe Chauvin,F-95302, Cergy-Pontoise Cedex, France}
\affil[2] {Institute for Biocomputation and Physics of Complex Systems (BIFI), University of Zaragoza, Spain}
\affil[3] {Department of Theoretical Physics, Faculty of Sciences, University of Zaragoza, Spain}
\affil[4] {ISI Foundation, Turin, Italy}

\date{}
\begin{document}
\maketitle

\begin{abstract}

1. Nestedness is a property of interaction networks widely observed in natural mutualistic communities, like plant-pollinators or plant-seed dispersers, among other systems. A perfectly nested network is characterized by the peculiarity that the interactions of any node form a subset of the interactions of all nodes with higher degree. Despite a widespread interest on this pattern, no general consensus exists on how to measure it. Instead, several metrics aiming at quantifying nestedness, based on different but not necessarily independent properties of the networks, coexist in the literature blurring the comparison between ecosystems. 

2. In this work we present a detailed critical study of the behavior of six popular nestedness metrics and the variants of two of them. In order to evaluate their performance, we compare the obtained values of the nestedness of a large set of real networks among them and against a maximum entropy and maximum likelihood null model. We also analyze the dependencies of each metrics on different network parameters as size, fill and eccentricity.

3. Our results point out, first, that the metrics do not rank the degree of nestedness of networks universally. Furthermore, several metrics show significant undesired dependencies on the network properties considered. The study of these dependencies allows us to understand some of the systematic shifts between the real values of nestedness and the average over the null model.

4. This paper intends to provide readers with a critical guide on how to measure nestedness patterns, by explaining the functioning of six standard metrics and two of its variants, and then disclosing its qualities and flaws. By doing so, we also aim to extend the application of the recently proposed null models based on maximum entropy to the still largely unexplored area of ecological networks.

5. Finally, to complement the guide, we provide a fully-documented repository named \emph{nullnest} which gathers the codes to produce the null model and calculate the nestedness index -both the real value and the null expectation- using the studied metrics. The repository contains, moreover, the main results of the null model applied to a large dataset of more than 200 bipartite networks.

\end{abstract}

\section{Introduction}

The characterization of mutualistic networks has been the ground of considerable debate during the last decades. These type of networks are represented as a graph that codifies mutually beneficial interactions, namely, the species of the network involved in these interactions naturally obtain a  benefit from them, even if their nature could be different. This is the case, for instance, of plant-pollinator communities where pollinators feed on flower's nectar while plants assure their reproduction. Moreover, as mutualistic interactions often take place only between species of different kind, they can therefore be represented by a {\em bipartite network}, characterized by two disjoint sets of vertices (or nodes) representing the species, with the edges (or links) joining only vertices of different kind, i.e., links connect species of the two branches of the bipartite graph.

The structure and dynamics of mutualistic networks have received increasing attention due, in particular, to the role that mutualism is assumed to play in the complexity-stability paradox as a stabilizer of large and complex communities~\cite{mccann2000diversity}. Indeed, it has generally been admitted that mutualistic interactions enhance stability by screening competition~\cite{bastolla2009architecture,thebault2010stability}, though this idea has recently been challenged~\cite{james2012disentangling,staniczenko2013ghost,gracia2018joint}. 
Moreover, the observation of natural ecosystems has revealed that in a vast majority of cases, mutualistic interactions are not uniformly distributed. Instead, the species interact in a very particular way, leading to a network structure called \emph{nestedness}~\cite{bascompte2003nested,bascompte2006asymmetric,fortuna2010nestedness}.

\begin{figure*}[h!]
\center
\includegraphics[width=0.8\textwidth]{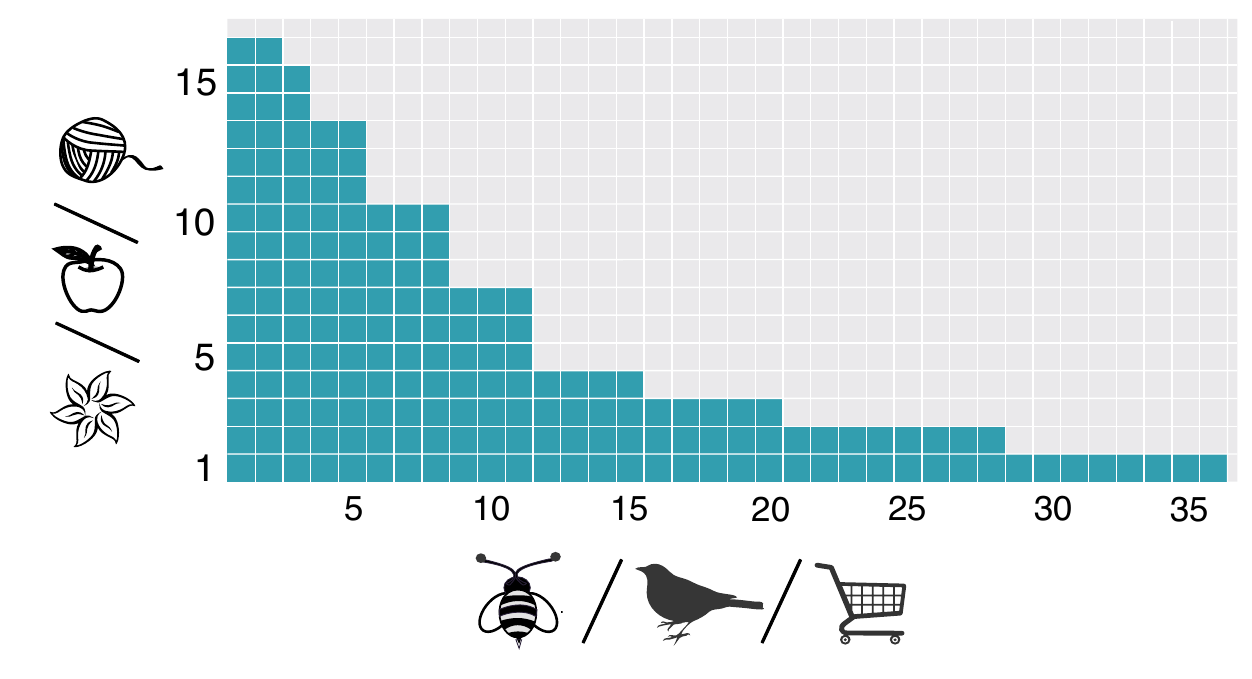}
\caption{{\bf  Scheme of a perfectly nested bipartite matrix.}
The region in green represents the "1s" in the bi-adjacency matrix, and the one in light grey the "0s". Nodes are ordered by increasing ranking (decreasing degree) from bottom to top and from left to right. Since the network is perfectly nested, the interactions of a given node are always a subset of the interactions of the nodes with smaller ranking. We portray three different possible types of paradigmatic mutualistic networks: flowering plants and its pollinators, fruit-producing plants and its seed-disperser birds, and sellers and buyers. The figure also shows how different nodes may have the same degree, leading to some degeneracy in the ordering.}
\label{fig:perfect_nestedness_matrix}
\end{figure*}

A network is said to be {\em perfectly nested} when the contacts of a species of a given degree are a subset of the contacts of all the species of larger degree, as illustrated in Fig.~\ref{fig:perfect_nestedness_matrix}. The system is then composed of generalist and specialist species in each  guild, the former interacting with a large amount of the possible counterparts and the latter only with generalists, in such a way that specialist-specialist interactions are mostly absent~\cite{bastolla2009architecture,thebault2010stability,okuyama2008network,olesen2007modularity}. As a consequence, when the nodes of one guild are ordered by decreasing (or increasing) degree, the nodes of the other guild appear automatically ordered in the same way, and the corresponding bi-adjacency matrix has all its non-zero elements on the same side of a curve called "isocline of perfect nestedness" (IPN), see Fig.~\ref{fig:continuous_ipn}. This ordering leads to the characteristic triangular shape~\cite{medan2007analysis} shown in Fig~\ref{fig:perfect_nestedness_matrix}. The arrangement of the matrix which reveals and maximizes its nestedness is usually referred to as a {\em maximally packed configuration}, and various methods to produce it can be found in the literature~\cite{rodriguez2006new,dominguez2015ranking,lin2018nestedness}. 

Although natural mutualistic ecosystems are not perfectly nested, the same general pattern with some fluctuations is found in a large variety of known ecosystems, that correspond to a wide range of different geographic and climatic conditions and involve very different species~\cite{bascompte2003nested}. Despite that nested patterns have mainly been observed and studied in ecology, systems displaying nested forms of interactions have also been reported for a variety of economic and social networks~\cite{hernandez2018trust,borge2017emergence}.

The astounding ubiquity of nestedness in natural systems called for the need of both having a good indicator to quantify it, and providing answers to challenging questions about the reasons behind such a generally observed pattern and its consequences. Regarding the latter problem, a recent work~\cite{payrato2019breaking} has shown analytically and numerically that nestedness is not an irreducible global property, but that it emerges instead from local properties of the network. In particular, it arises as an entropic consequence of the double heterogeneity observed in the degree sequences of natural ecosystems. Indeed, building a maximum entropy ensemble with the constraint that the observed degree sequence is kept on average such that this sequence is found with maximum probability in the ensemble, one finds that nestedness values of real networks do not significantly differ from the average of the random ensemble. Ulrich and Gotelli~\cite{ulrich2007null} and Jonhson et al.~\cite{jonhson2013factors} had also observed this fact when randomizing real networks but fixing the degree sequences exactly (the so-called FF model, see below). However, the poor statistics that results from such strong constraint limited the understanding of the extent of such observation.  All in all, the fact that nestedness is not an independent pattern but that it stems from the heterogeneity of both degree distributions does not invalidate its usefulness. Indeed, its global character makes it a helpful tool to detect heterogeneity in species' connectivity, by using a single parameter. Given that the observed degree distributions are highly heterogeneous, it is tempting to characterize them by the exponent of a power law fit, and several attempts were made in this sense~\cite{jordano2003invariant}. However, the typical small size of natural ecosystems makes such fit meaningless. Therefore, nestedness continues to be a useful and widely studied global pattern, and the aforementioned question of how to measure it remains an essential challenge.

Admittedly, as a result of the effort to quantify nestedness, a variety of {\em metrics} with their corresponding {\em nestedness indices} coexist in the ecological literature. However, since they are based on diverse but not necessarily independent properties of the nested networks, how to compare the degree of nestedness of different ecosystems remains unclear. The situation recalls the well-known history of the definition of temperature in  Thermodynamics. Initially defined operationally, i.e., by listing the protocol to measure it, the obtained temperature values suffered from  the flaw that they  depended on the thermometer  used. This problem was solved  by the theoretical definition of the temperature based on the Second Principle of Clausius, and finally the notion of temperature was completely understood by the microscopic approach of Statistical Physics introduced by Boltzmann and Gibbs. Interestingly, the first  metrics of nestedness defined by Atmar and Patterson was called \emph{temperature}~\cite{atmar1993measure}. This initial proposal was followed by a long struggle to find the best index to measure nestedness, with the development of various approaches ranging from algorithmic procedures to analytical methods. A review of the early nestedness indices by Ulrich and Gotelli was published in 2009~\cite{ulrich2009consumer}, while a more recent review by Mariani et al.~\cite{mariani2019nestedness} provides a very detailed and updated summary of the most common nestedness metrics.

Nonetheless, the metrics defined to quantify nestedness suffer from a critical drawback: as they are strongly dependent on different network parameters (like size, fill, etc), the comparison among ecosystems is difficult, even in the case where the same metrics (the same thermometer) is used to measure all the systems. These problems have been reported by several authors, notably on the occasion of the introduction of each new index and/or package devoted to correct some of the shortcomings of  previously existing ones~\cite{rodriguez2006new,almeida2008consistent,dormann2009indices,burgos2009understanding,galeano2009weighted}. Still, these works mainly focus on the dependence on the size and the density of links of the network of a few metrics, leaving aside other important nestedness indices as well as the interdependencies among network parameters. 
  
In order to overcome the aforementioned difficulties when measuring and comparing the nestedness of different networks, the standard procedure is to contrast the nestedness value of a given real network with that of a {\em null model}, both calculated using the same metrics. In particular, a null model is an ensemble of networks obtained by the randomization of the natural system under study, imposing some constraints. Different constraints lead, then, to different null models of the same real network~\cite{ulrich2007null,payrato2019breaking}. In a majority of cases, such constraints are enforced algorithmically. In particular, a very popular choice is the \textit{fixed-fixed} null model (FF herein), where the degree sequences are strictly kept and null networks are produced through numerical randomizing procedures~\cite{gotelli2001swap}. On the other hand, the family of null models based on maximum-entropy ensembles precludes the algorithmic randomization~\cite{squartini2011analytical}, and they have just recently begun to be applied to the study of ecological networks~\cite{payrato2019breaking}. However, while the majority of nestedness metrics have been tested for algorithmically-based null models~\cite{ulrich2007null,almeida2008consistent}, their behavior in maximum-entropy ensembles is still largely unexplored.

In this work, we focus on the problem of measuring nestedness by presenting a comparative study of the behavior of six nestedness metrics most of which are commonly included in popular packages and cited in the literature. Our purpose is two-fold: first, we aim to test the performance of these metrics under the maximum-entropy null model recently used in~\cite{payrato2019breaking}, and secondly, we intend to critically assess the functioning of each metrics by analyzing its dependencies with network parameters. By doing this we mean to, first, fill a gap in the literature concerning null models, and second, to provide a practical guide of the advantages and disadvantages of each nestedness metrics. To this end, we study the nestedness metrics using the following procedure: for each of the 199 real bipartite networks of our dataset, we build the corresponding maximum entropy and maximum likelihood ensemble that preserves on average the observed degree sequence~\cite{squartini2011analytical,payrato2019breaking}. We then measure nestedness in the ensemble built for each of the real networks according to each of the metrics and we compare the results with the corresponding nestedness value of the observed network. Secondly, we perform various statistical analyses to determine the relation of each metrics with network properties, like size, fill, degree degeneracy, etc. 

The rest of the paper is organized as follows: in section~\ref{Metrics and models}, we briefly describe each of the metrics scrutinized in this work. Our results are presented in section~\ref{Results} together with a thorough comparison among the different metrics. More details about the calculations done and the methods employed are provided in section~\ref{Methods}, which is followed by a general discussion in section~\ref{Conclusions}. The metrics are reviewed and analyzed independently, so that a reader interested in a particular metrics may directly browse through its dedicated subsections. Moreover, as explained in section~\ref{Methods}, a working package called \emph{nullnest} is provided, which contains: a) the probability matrices of contacts among the species allowing to reconstruct the maximum entropy, maximum likelihood null model of the studied real networks, b) the documented program that produces this probability matrix to be used with other real matrices, c) the documented programs that calculate all the indices on the real and randomized matrices.

\section{Metrics and models}
\label{Metrics and models}
\subsection{The studied metrics}
\label{Metrics}

We briefly describe here the principal characteristics of the indices used in this work in order to quantify nestedness. The technical details on how each metrics has been numerically implemented can be found in section~\ref{Methods}.
\begin{itemize}

\item  {\bf The {\em Atmar and Patterson temperature} ($T_{AP}$)}~\cite{atmar1993measure}. This nestedness metrics is based on the idea of quantifying the deviations of a real matrix from a perfectly nested matrix by measuring the distance of the misplaced interactions from the IPN curve. Various implementations of this metrics can be found in the literature~\cite{rodriguez2006new,atmar1995nestedness,guimaraes2006improving}, however the most popular nowadays is BINMATNEST, developed by Rodríguez-Gironés and Santamaría~\cite{rodriguez2006new}. 

The mathematical basis of this metrics~\cite{medan2007analysis} relies on the  mapping of the maximally packed version of a $m \times n $ bipartite adjacency matrix (as shown in Fig~\ref{fig:perfect_nestedness_matrix}) into a continuous rectangle, leading to  the analytic expression of the IPN in terms of  two continuous variables $a \in [0,n]$ and $p \in[0,m]$, which constitute the continuous approximation to the discrete labels of the columns and rows of the bipartite adjacency matrix, respectively. This approximation  is expected to be correct in the limit of very large systems. Then, the non zero elements of the adjacency matrix correspond, in the rectangular surface of size $m \times n$, to an area proportional to the density of contacts $\phi = E/(m \times n) $, where $E$ is the total number of edges. This area may be assumed to be coloured and so the empty area represents the amount of zero elements of the adjacency matrix. The IPN can be determined as a function of  $m$, $n$ and $\phi$~\cite{medan2007analysis}. The practical implementation of this metrics in BINMATNEST rescales the axis into the unit square.

Because real systems are not perfectly nested, the $T_{AP}$ measures the distance, along the diagonal of the unit square, of the misplaced points (presence or  absence of a contact above or below the IPN respectively)~\cite{atmar1993measure}. In particular, Rodríguez-Gironés and Santamaría~\cite{rodriguez2006new} proposed to quantify the \textit{unexpectedness} of a given interaction by the following function: 

\begin{equation}
u_{ij} \; = \; \left ( \frac{d_{ij}}{D_{ij}} \right) ^2 \, ,
\label{eq:temperature}
\end{equation}

where $d_{ij}$ and $D_{ij}$ correspond, respectively, to the distance between the unexpected interaction and the IPN in the first case and to the total length of the diagonal in the second (see Fig.~\ref{fig:continuous_ipn}b for an illustration). The final temperature is then calculated as follows:
\begin{equation}
T_{AP} \; = \; \frac{100}{U_{max} \cdot n \cdot m} \sum u_{ij} \, ,
\label{eq:temperature2}
\end{equation}

where the sum runs over all the unexpected interactions and $U_{max}$ is a constant given by Atmar and Patterson~\cite{atmar1993measure}. Accordingly, the $T_{AP}$ will be large if there are several $'1s'$ and $'0s'$  on the wrong side of the IPN. It will be even larger if those misplaced points are located far from the IPN. Then, the lower the measure of the $T_{AP}$ of a given system, the more nested it is. 
 
\begin{figure*}[h!]
\center
\includegraphics[width=\columnwidth]{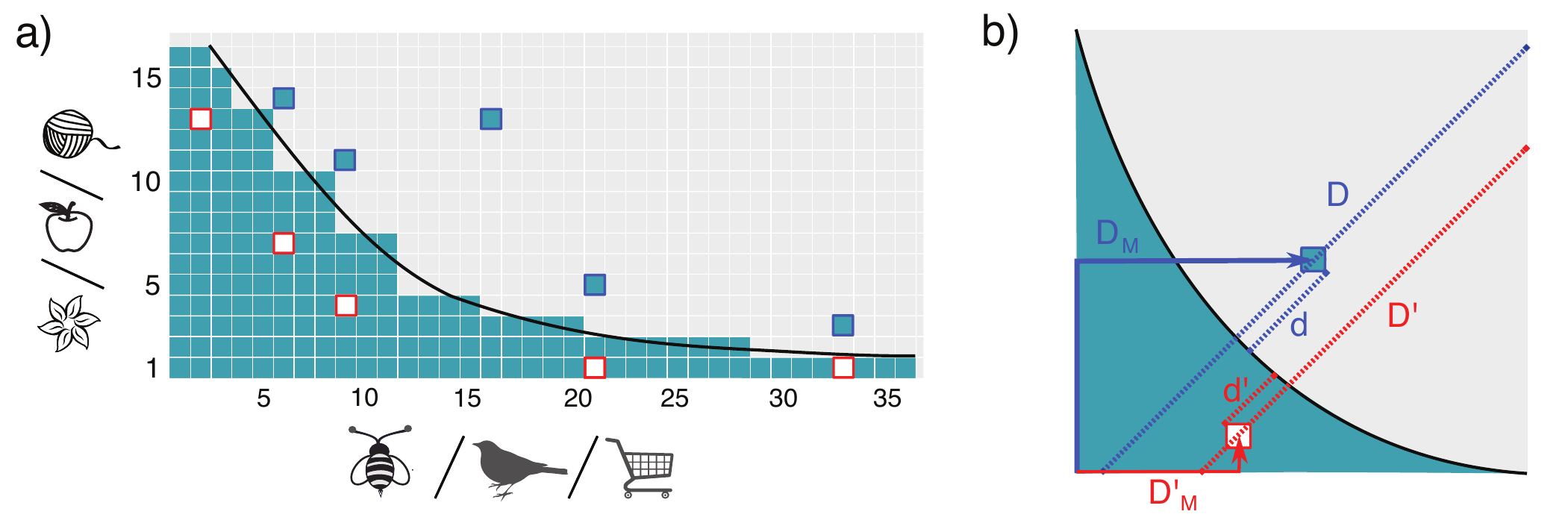}
\caption{{\bf Continuous approximation of a maximally packed nested matrix.}
Figure \textit{a)} shows the scheme of a non-perfectly nested network. Green squares depict the interactions and the black curve represents the {\em Isocline of perfect nestedness} (IPN) which separates, in an ideally perfect nested matrix, the region where all the nodes are connected from the region with no interactions. The unexpected interactions above the IPN are highlighted in blue, while absent interactions below the IPN are highlighted in red. Figure \textit{b)} represents the mapping of a matrix into the unit square. The black curve corresponds to the IPN, and unexpectedly present (absent) interactions are highlighted in blue (red). The figure shows two different kind of distances that may be used for measuring nestedness: $D$, $D'$, $d$ and $d'$ (traced in dashed lines) are used in the calculation of the temperature, while $D_M$ and $D'_M$ (represented by solid arrows) are used by the nestedness index based on Manhattan distance (NMD).}
\label{fig:continuous_ipn}
\end{figure*}

\item {\bf The {\em nestedness index based on Manhattan distance} (NMD)}~\cite{corso2008new}.
This metrics follows the same idea as the $T_{AP}$ metrics, in the sense that it counts the number of unexpected presences or absences with respect to a nested matrix of the same characteristics (size and fill) as the studied matrix, when both are brought to their maximally packed form. Again, it introduces a mapping of the matrix into the unit square. On this rescaled continuous approximation, it measures the distance to the corner of the matrix where the nested core is expected (see Fig.~\ref{fig:continuous_ipn}b for an example). Distances are measured in terms of the Manhattan distance, which means that the distance between a rescaled element $b_{i,j}$ of the matrix and the origin is $d_{i,j}=x_i+y_i$. The nestedness index is then given by:

\begin{equation}
\tau=\frac{d-d_{nest}}{d_{rand}-d_{nest}} \, ,
\label{eq:manhatan}
\end{equation}

where $d$ is the sum over all the elements' distances $d=\sum d_{i,j}$ and $d_{nest}$ represents an analogous sum but over the maximally nested matrix. Their difference is then normalized by the maximum difference in average distances between a null model and the perfectly nested matrix. In this way $0\leq \tau \leq 1$, and the smaller $\tau$ the more nested the system is. Here we used the implementation of the popular \textit{bipartite} package~\cite{dormann2009indices} where the null model used to calculate $d_{rand}$ keeps constant size and fill (see Section~\ref{Methods} for details). 

\item {\bf The {\em nestedness metrics based on overlap and decreasing fill} (NODF)}~\cite{almeida2008consistent}. This index measures the average percentage of shared contacts between pairs of rows which present a decreasing degree ordering in their maximally packed form (idem for columns). Therefore, this metrics separately informs on the  contribution of  rows and of columns to the observed nestedness. It should be noted that the overlaps between all the possible pairs of  rows (columns) are only taken into account if the considered pair is ordered in decreasing degree, otherwise it assigns a null value to the overlap. For this metrics, the higher the NODF index, the more nested the system is.

NODF has the advantage that it can be calculated not only algorithmically but also by using a closed mathematical expression in terms of the elements of the bi-adjacency matrix, which allows for analytic studies~\cite{payrato2019breaking}.
It correctly assigns a very low nestedness value to modular networks (because, in general, elements within the same block have similar degree), but it may give a false negative (a low value) in the case of a nested network with multiple rows (columns) with the same degree~\cite{staniczenko2013ghost}. Unfortunately, this situation is quite common for mutualist ecosystems which are in general very sparse and often eccentric, with typically much more animal species than plant species, leading to a non negligible degree degeneracy. For this reason, a variant of this metrics called \textbf{\em{stable-NODF}} has recently been proposed by Mariani et al.~\cite{mariani2019nestedness}, which does not incorporate the decreasing fill term and hence does not penalize the degree repetition. Thus, this metrics measures solely the number of shared partners among pairs of rows and columns. Its analytical expression can be found in~\cite{payrato2019breaking}. 

\item {\bf The {\em Brualdi and Sanderson discrepancy}}~\cite{brualdi1999nested}. Starting from the real matrix in its maximally packed state, this metrics measures the number of misplaced absences or presences of contacts, called \textit{discrepancies}, that should be \textquoteleft corrected\textquoteright ~in order to produce a perfectly nested matrix with equal size and fill. Since this metrics is based on the comparison of the real matrix with a perfectly nested one of the same parameters ($n$, $m$ and $\phi$), it is independent of a particular null model. However, since there may be some ambiguity on the perfect configuration, the result depends on the chosen one. Therefore the best approach would involve averaging over the different initial maximally packed configurations of the observed matrix, yet this procedure is very demanding numerically  and we do not implement it in this work.
 
Furthermore, given that the number of possible discrepant links is directly proportional to the total number of links, the result greatly depends on the network's fill. To hinder the results from such dependency, we normalized the discrepancy by the total number of links, as suggested by Greve and Chown~\cite{greve2006endemicity}. As it was the case of the temperature $T$, the  definition of this metrics implies that the lower the value of the index, the more ordered the system is.

\item {\bf The {\em nesting index based on network's robustness} (NIR)}~\cite{burgos2009understanding}. 
NIR metrics is based on the notion of the robustness of a network, that is, the capacity of the system  to remain connected when subject to node removal~\cite{burgos2007nestedness,memmott2004tolerance}. This metrics uses two extreme node removal procedures, or {\em attack strategies}, whose outcomes reveal the amount of nestedness of the network. On the one hand the nodes of one guild are removed in {\em decreasing degree order} (DDR strategy), and of the other in {\em increasing degree order} (IDR strategy). The fraction of species of the other guild that still keeps contacts (survive) as the counterparts are removed leads to the  {\em Attack tolerance curve} (ATC).

Once the attack strategy is fixed, the  ATC depends on the  degree of nestedness. Fig~\ref{fig:atc} illustrates three different typical behaviors of the  ATC for each strategy, when the procedure is applied on a perfectly nested network, on a real network and on a null model with the same size and fill. The DDR strategy better reveals the differences of structure of the three networks. 

\begin{figure*}[h!]
\center
\includegraphics[width=0.6\textwidth]{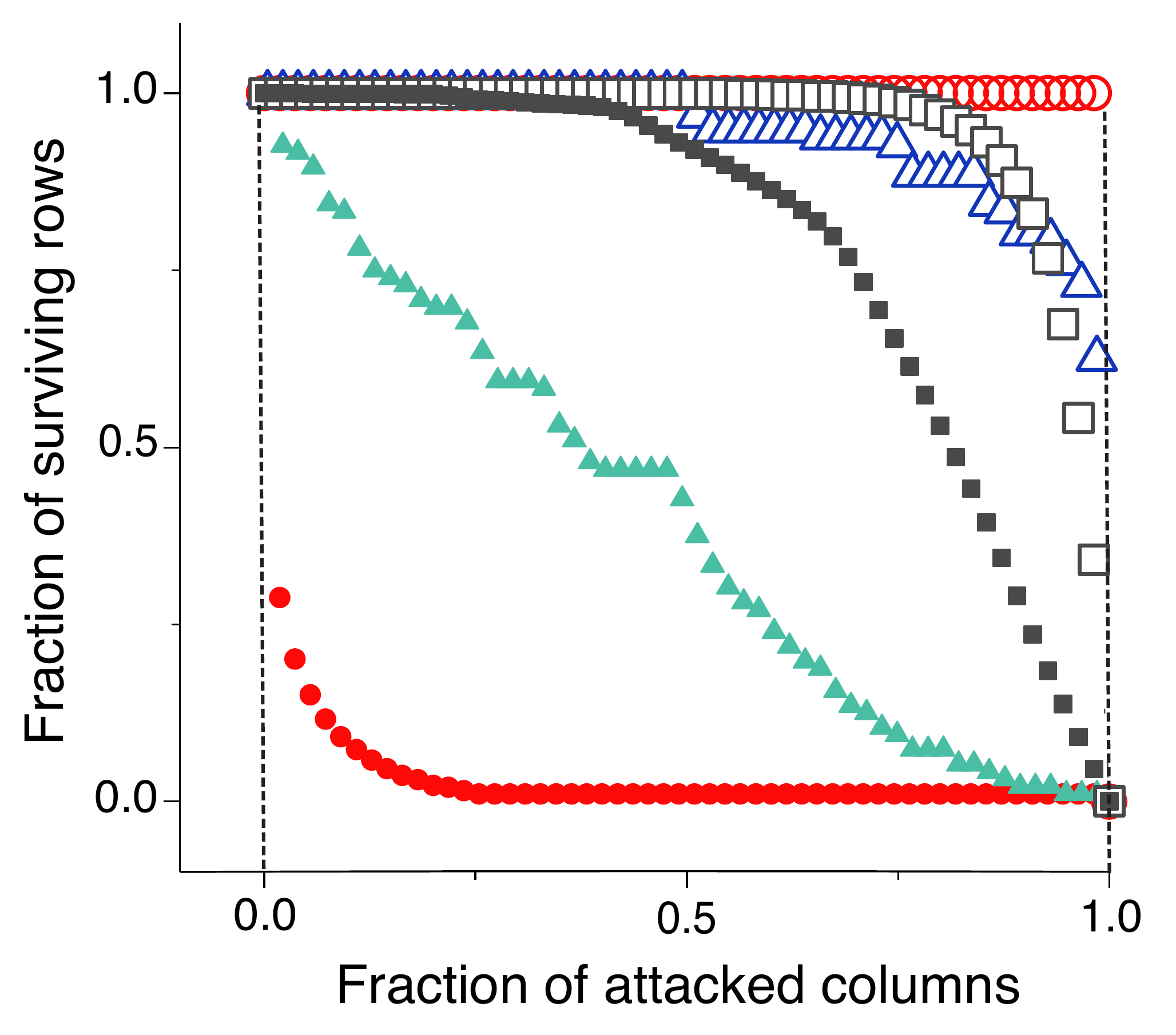}
\caption{{\bf Attack Tolerance Curves for three different networks having the same parameters.}
Triangles correspond to  the real mutualistic ecosystem of Clements and Long~\cite{clements1923experimental}, squares to a randomization of this system and circles to an artificial perfectly nested network with the same parameters (size and number of links). Open and full symbols  correspond to the IDR and DDR attack strategies, respectively.}
\label{fig:atc}
\end{figure*}

It can be easily shown~\cite{burgos2009understanding} that, for the perfectly nested network, the area under the ATC is $R_{IDR}=1$ for IDR strategy, while it is $R_{DDR}= \phi$, for the DDR.  Thus, this index is normalized by the area between these extreme curves, which is maximum for a nested network. Moreover, the area is minimum for a random network, while for the real networks the area lies between these two extremes. Therefore, the contribution to the nestedness coefficient of rows or columns is defined as:

\begin{equation}
NIR=\frac{R_{IDR}-R_{DDR}}{1-\phi},
\label{nesting_coeff}
\end{equation}

which measures, like NODF, the contribution to nestedness of rows and columns, separately. NIR looses sensitivity as the density of links increases, which is not a problem for ecosystems that are, in general, very sparse. Finally, this index may in principle slightly depend on the chosen matrix ordering with respect to the degrees of the guild being suppressed. As such order is not unique due to degree degeneracy, in our implementation we have  averaged over a set of equivalently ordered matrices.

\item {\bf The {\em spectral radius}}~\cite{staniczenko2013ghost} . A recent article~\cite{staniczenko2013ghost} proposed a nestedness metrics based on  the spectral properties of double nested graphs. It has been proven~\cite{bell2008graphs,bhattacharya2008first} that among all the connected bipartite graphs of $n + m$ nodes and with $E$ edges represented by its adjacency matrix $\cal{A}$, the one yielding the largest spectral radius is a perfectly nested graph. Staniczenko et al.~\cite{staniczenko2013ghost} then showed that this theorem entails a statistical behaviour by which more nested graphs \textit{tend} to have as well a larger spectral radius (though this relation is not monotonous). Interestingly, since this metrics involves diagonalizing a symmetric matrix, it is independent of the ordering of the matrix and does not suffer from the ambiguities of finding the maximally packed form of the bipartite matrix as the previous ones. 

Despite the mentioned qualities, an important drawback of the spectral radius is that it is not normalized. Remarkably, the theorem previously cited requires the size of the matrix and the number of links to be fixed, a condition that should be considered in order to set a benchmark against which to compare the nestedness index of a real network. Although the FF null model respects the required hypothesis, it is well known that it leads to a poor statistics due to the relatively few number of networks matching these constraints in finite (and small) systems. For these reasons, we propose to normalize the spectral radius obtained for each system with that of the perfectly nestedness matrix having the same size and fill (see Section~\ref{Methods} for details), as was already suggested by Staniczenko et al. That is, if $\rho$ represents the spectral radius of a real network and $\rho_{max}$ the spectral radius of a perfectly nested graph with the same size and fill, the normalized index $\rho_{norm}$ is given by the expression:

\begin{equation}
\rho_{norm} = 100 \, \frac{\rho}{\rho_{max}} \label{eq:spectral_radius}
\end{equation} 

In the next sections, we study both versions of this metrics: the (not normalized) original one along with the normalized modification given by Eq.~\ref{eq:spectral_radius}.

\end{itemize}

\subsection{The maximum entropy-maximum likelihood realization of the FF null model}
\label{Null model}

We use a null model for bipartite networks that constrains the degree sequence of each guild, so that they are kept only \textit{on average}. That is, at variance with the previously studied FF null model, where the real degree sequences are enforced \textit{strictly}~\cite{ulrich2007null}, in our case, the degree sequences of a sampled null network may slightly vary from the real ones, with the restriction that the average degree sequences are kept. More precisely, the statistical ensemble of networks generated by our null model is required to fulfill two conditions:

\begin{enumerate}[(i)]
\item It is a maximum-entropy ensemble, meaning that minimal information about the links of each network of the ensemble is conveyed by the randomizing process. The unique requirement is imposed to the average degree sequences of each guild in the ensemble, which must be equal to the observed degree sequences of the empirical network under study, leading to a constrained maximization of the entropy.
\item The degree sequence of the empirical network must be found in the generated ensemble with maximum probability.
\end{enumerate}

These two requirements warrant that the resulting ensemble is maximally disordered while enforcing the average degree sequences in the ensemble to be equal to the  degree sequences of the corresponding real network, which in turn are the most probable degree sequences for each guild. This randomizing framework was first developed by Squartini and Garlaschelli~\cite{squartini2011analytical}, then extended to bipartite networks by Saracco et al.~\cite{saracco2015randomizing}. We also applied it to the study of the emergence of nestedness in ecological networks in~\cite{payrato2019breaking}, where the interested reader will find a more detailed description of the null model. 

A conceptual difference between the present and previous null models is that the resulting ensemble of networks is treated from a statistical physics perspective, that is, by using the probability of appearance of each network in the ensemble. When imposing that the entropy shall be maximized, one encounters the Exponential Random Graph model. It is important to recall that this optimization of the entropy is constrained, since we aim to maintain each guild's degree sequences on average, which involves the introduction of some undetermined Lagrange Multipliers. Such indeterminacy can be solved by requiring that the most probable degree sequences in the ensemble must be those of the real bipartite network. This ensures that the statistical ensemble results in a randomization of the real network, while enforcing the average degree sequences in the ensemble to be equal to the real ones.
 
Furthermore, this double maximization yields an \textit{analytic expression}    for the probability of interaction between any two nodes from different guilds in the statistical ensemble. Remarkably, this probability only depends on the Lagrange Multipliers. These variables, in turn, can be determined by computationally solving the optimization problem given in \textit{(ii)}. We have used two different numerical methods to find the global optimal Lagrange Multipliers for each empirical network: \textit{(a)} a global searching algorithm based on simulated annealing, and \textit{(b)} a local optimization method repeated over a variety of initial conditions. The technical details of the implementation of these techniques can be found in~\cite{payrato2019breaking}. We performed both analysis for each of our real networks and verified that the results agree, which is a strong indication that the global maximum has been found.

The present null model overcomes many of the statistical bias exhibited by the FF null model. On the one hand, loosing the constraint of preserving exactly the degree sequences yields to an expansion of the statistical ensemble. On the contrary, with the FF null model the number of null networks compatible with the constraints becomes scarce whenever the real network we aim to randomize is significantly small, dense or nested. On the other hand, the double maximization of the entropy and the likelihood produces a maximally disordered statistical ensemble while constraining certain information extracted from real systems, in our case the empirical degree sequences. Garlaschelli and Loffredo showed~\cite{garlaschelli2008maximum} that such construction is statistically non-biased. Finally, the fact that our null model provides an analytic expression for the probability of interaction between species, results in the computational generation of null networks being fast, efficient and demanding few numerical resources. Instead, the FF null model relies on an algorithmic procedure that can easily become frustrated, thus slowing down the production of null networks and being computationally demanding. All these advantages, together with the conceptual basis argued in~\cite{payrato2019breaking}, lead us to choose this null model to assess the performance of a variety of nestedness metrics. 

The reader will find a summary on how we produced the sampling of null networks for each real network in section~\ref{Methods}. Moreover, the documented \textit{nullnest} github repository we supply includes the probability interaction matrices that allow to generate the null model for each of the studied real networks. Likewise, we provide the codes that yield the null model for any other network.

\section{Results}
\label{Results}

\subsection{Significance of nestedness of empirical networks.}
\label{Results:null_model}
We have measured the nestedness of 191 empirical ecological networks extracted from the \textit{Web of Life}~\cite{weboflife} as well as 8 economic networks which represent the trading interactions between the buyers and the sellers of two different fish markets~\cite{hernandez2018trust}. To compare the average nestedness over the ensemble with that corresponding to empirical networks, we have used the six metrics described above plus two variations (the stable NODF and the normalized spectral radius). The details about the networks dataset can be found in section~\ref{Methods}. 

As it has been shown analytically and numerically (using NODF and the spectral radius) in~\cite{payrato2019breaking}, the nested structure of mutualistic networks is a consequence of the double heterogeneity in the degree sequence which results from entropic effects. In order to investigate if the other popular indices are able to reveal this dependence of the nestedness on the  degree distributions, we built a null model for each real network, as explained in~\ref{Null model}, and we compared the nestedness of each real network with that of the average over the ensemble. For each of the studied metrics, the average value of nestedness over the randomized ensemble has been obtained by numerically sampling the ensemble, using the probability interaction matrix described in section~\ref{Methods_null}.

\begin{figure}[h!]
\includegraphics[width=\textwidth]{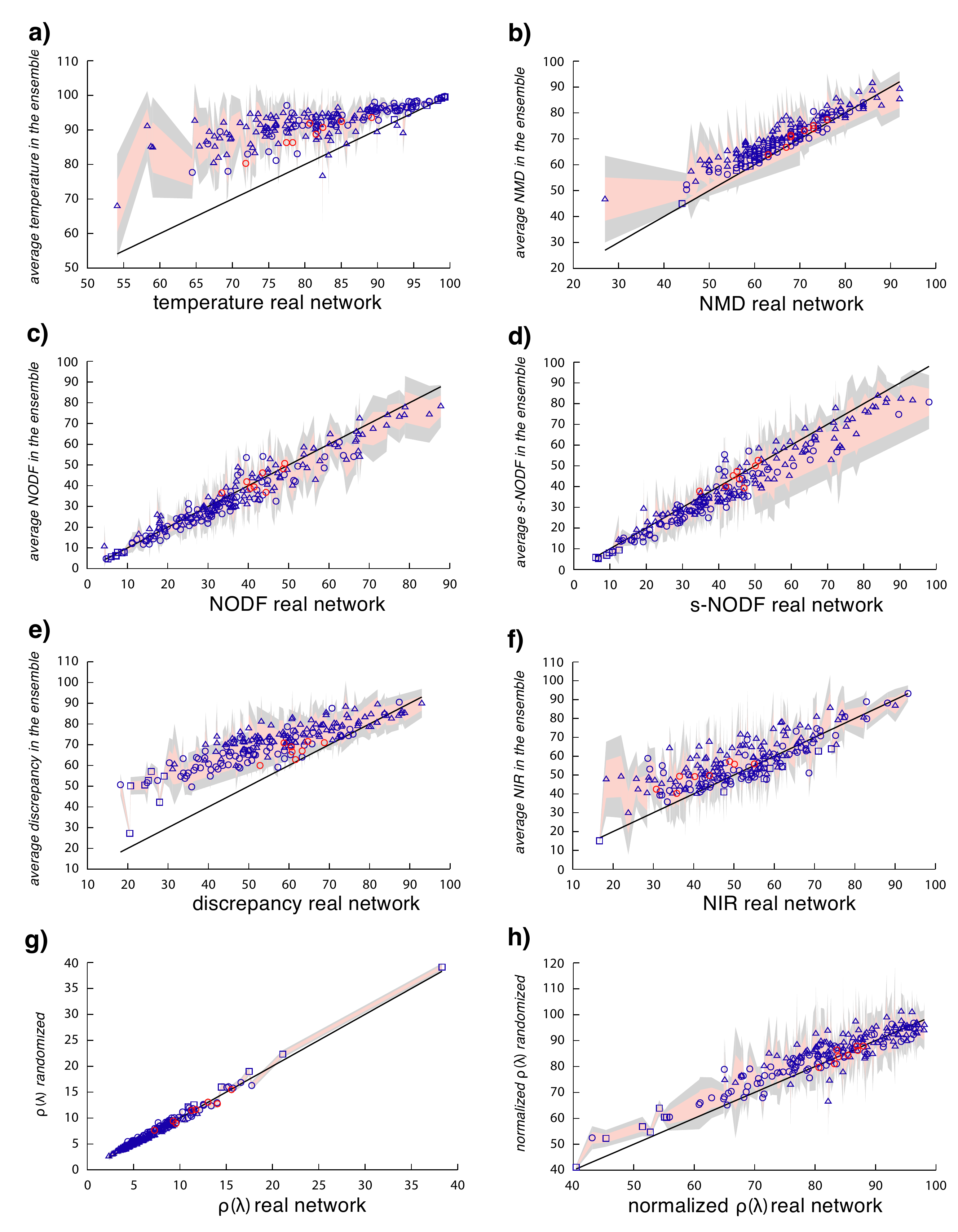}
\caption{{\bf Significance of the nestedness of real networks.} The figure shows the empirical measure of nestedness against the average value of nestedness in the generated statistical ensemble for the 199 empirical networks in our dataset. The different panels correspond to different metrics: (a) temperature, (b) NMD, (c) NODF, (d) stable-NODF, (e) discrepancy, (f) NIR, (g) spectral radius and (h) normalized spectral radius. The shadowed areas represent one (salmon color) and two (light gray color) standard deviations of the mean. The black line depicts the identity curve. Triangle symbols stand for small networks (less than 50 nodes), circles for medium size networks (more than 50 nodes and less than 410) and squares for large networks (more than 410 nodes). Ecological networks are colored in blue, economic networks in red.}
\label{fig:null_model}
\end{figure}

For the sake of clarity and to homogenize the reading of diverse figures, we have transformed the definition of the temperature, the NMD and the discrepancy indices so that the larger the index the more nested the system is. We have also rescaled these indices so that they vary between 0 and 100. These modifications read as follows:

\begin{eqnarray}
T=100-T_{AP} \, , \label{eq:redef_T}\\  
NMD= 100 \, \left( 1-\tau \right) \, , \label{eq:redef_NMD}\\
\Delta'= 100 \, \left( 1-\frac{\Delta}{E} \right). \label{eq:redef_dis}\\
\end{eqnarray}

where $\Delta$  is the discrepancy index described in~\ref{Metrics}, and E the total number of edges in the network.
Figure \ref{fig:null_model} shows the nestedness measured over the ensemble versus the nestedness of the corresponding real network.  Consistently with the results obtained in~\cite{payrato2019breaking} using NODF and the spectral radius, NIR and NMD also show that the nestedness values of the empirical networks are statistically equivalent to the average of the corresponding randomized ensemble. This leads to the conclusion that the observed nestedness measured by these indices is not significant. On the contrary, the discrepancy and temperature indices show a clear bias, with an important fraction of the real networks being {\em less} nested than the random average.

\subsection{Influence of network properties on the behavior of the different metrics}

\begin{figure}[h!]
\center
\includegraphics[width=\textwidth]{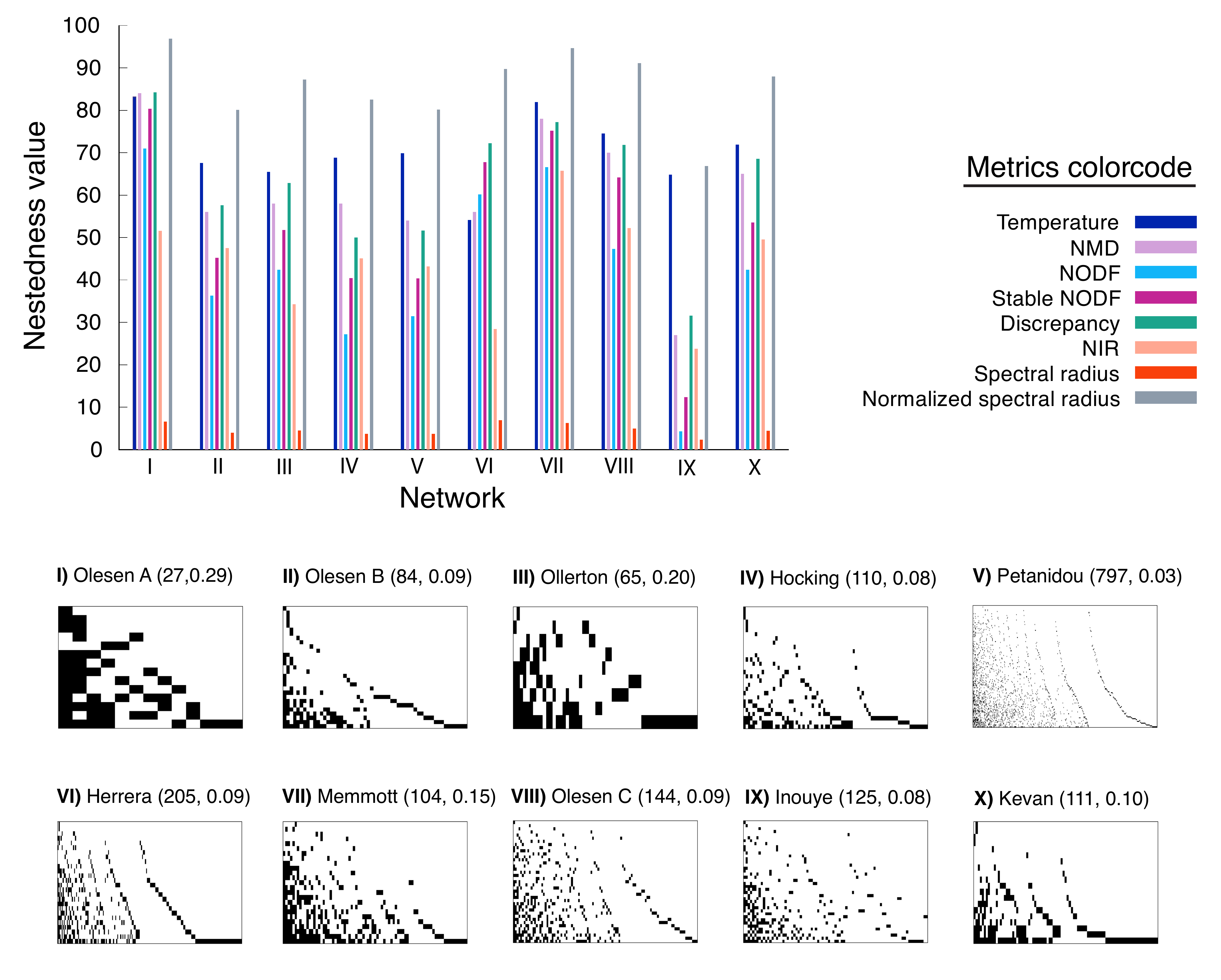}
\caption{\textbf{Comparison among nestedness indices.} The histogram on the top of the figure shows how eight different metrics measure the nestedness of several different networks. Each network, indexed I to X, is represented in the bottom of the figure by its bi-adjacency matrix ordered by decreasing degree, with the interactions among species represented by black pixels. All networks represent plant-pollinator mutualistic communities extracted from the Web of life dataset~\cite{weboflife}. Each network is labeled with the name of the first author of the corresponding reference, followed within brackets by, first, its total number of species (number of plants plus number of animals), and second, its density of links.}
\label{fig:all_metrics}
\end{figure}

The results presented in Fig.~\ref{fig:null_model} reveal that the metrics studied behave in different ways under the same null model, showing distinct levels of fluctuations and sometimes a systematic bias, as it is the case for the discrepancy and temperature indices. This finding suggests that the different algorithms implemented by each metrics may eventually translate into non-equivalent nestedness measures. We explore further this situation in Fig.~\ref{fig:all_metrics}, where we compare the values of nestedness obtained for a group of mutualistic networks when measured using each of the metrics. As it can be observed, for the same dataset not only the value of nestedness itself but also the ranking of the networks according to their degree of nestedness is strongly metrics' dependent.

Ideally, as it has been recalled by several authors~\cite{staniczenko2013ghost,ulrich2009consumer,almeida2008consistent}, a well-behaved nestedness metrics ought to be independent of the particular network parameters and, furthermore, rank the degree of nestedness of a given set of networks \emph{universally}. The results discussed above put in evidence that the second condition is not always true. Regarding the first requirement, we next explore more carefully how the nestedness values given by each metrics depend on the network parameters. In particular, since the networks of the dataset cover a wide range of parameter values (see Fig.~\ref{fig:all_metrics} for an example), we analyze the effects of three characteristic network properties: size, density of links and eccentricity. These quantities are defined as follows:

\begin{align}
\text{size} \, \equiv s \, = \, n + m \, , \\
\text{density of links} \, \equiv \phi \, = \, \frac{E}{n+m} \, , \\
\text{eccentricity} \, \equiv \epsilon \, = \, \left | \frac{n-m}{n+m} \right |  \, , \\
\label{eq:ntw_parameters}
\end{align}

where, as before, $n$ and $m$ are, respectively, the number of rows and columns of the bi-adjacency matrix, while $E$ is the total number of links. The eccentricity quantifies the difference between the number of nodes of the two guilds, or in other words, the deviation from a square-shaped bi-adjacency matrix. Indeed, $\epsilon = 0$ for a square matrix and $\epsilon \to 1$ when one of the guilds is much larger than the other. Interestingly, most of the large ecological networks observed show more columns (animal species) than rows (plant species), with a frequent ratio of 1 to 3. This observation, though, cannot be generalized to all mutualistic networks, specially to small networks (which can be much more eccentric) or to non ecological systems.

Additionally, we study the dependence of nestedness on a fourth parameter, the degree degeneracy. In particular, a perfect nested matrix with an arbitrary $\phi$ might have several species of each guild with the same degree.  We measure this quantity as:

\begin{equation}
\text{degeneracy in degrees} \, \equiv \, g= \frac{\text{number of species with the same degree}}{n + m} \, .
\end{equation}

The study of this parameter remains a special case, since the known connection between the nested patterns and the degree sequences entails that a certain dependency with the degree degeneracy is in fact expected~\cite{payrato2019breaking,jonhson2013factors}. All in all, we analyze its influence given that each metrics deals with degree degeneracy in a different way.

\begin{figure}[h!]
\center
\includegraphics[width=\textwidth]{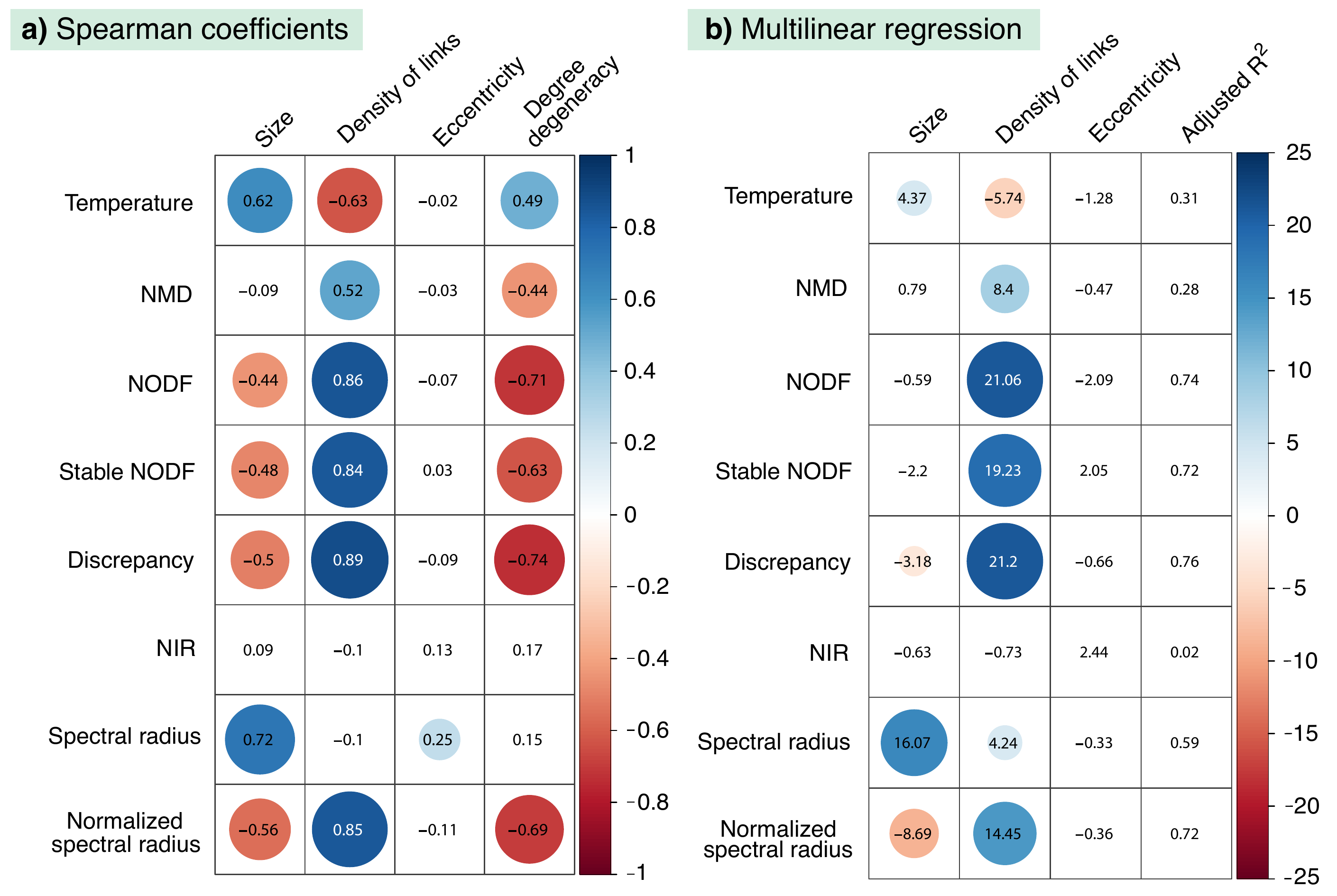}
\caption{\textbf{Dependency of nestedness metrics on network parameters.} The left panel, \textbf{a)}, shows the Spearman correlation factor between the networks parameters (columns) and the eight nestedness metrics under study (rows). The numbers represent the value of the Spearman rank coefficient for each corresponding pair of nestedness value and network parameter. Only those coefficients that are statistically significant ($p$-value $<0.01$) are highlighted by a colored circle, being the size and the color of the circle proportional to the coefficient. The right panel, \textbf{b)}, summarizes the results of the multi-linear fit detailed in Eq.~\ref{eq:multilinear_fit}. Each row corresponds to a different nestedness metrics. The first column from the right shows the adjusted coefficient of multiple determination (\textit{adjusted} $R^2$). The other three columns show the \textit{t-ratio} of the regression coefficient corresponding to each explanatory variable (as labelled by the column name). Only those coefficients that are statistically significant ($p$-value $<0.01$) are highlighted by a colored circle, being the size and the color of the circle proportional to its $t$-ratio.}
\label{fig:correlations}
\end{figure}

In order to quantify the dependencies discussed above we have performed a two-fold analysis. First, we have calculated the  Spearman's rank correlation between the nestedness index given by each metrics and the different network parameters. This coefficient allows to assess the relation between both variables without assuming a linear behavior. Fig.~\ref{fig:correlations}a summarizes the result of the analysis, showing the Spearman coefficient along with its statistical significance for all pairs of nestedness values and network parameters (see the Methods section for the details on the numerical calculation). Secondly, we have performed a multi-linear regression. In particular, we have taken the nestedness values obtained by each metrics as the dependent variable while the network parameters behave as the explanatory variables. Importantly, in this second analysis we do not consider the effect of the degree degeneracy, since we are mainly interested on the dependence on parameters that should \emph{not}, in principle, determine nestedness. The linear function we have fitted has the following standard form:
\begin{equation}
\nu_j \, = \, \beta_{0,j} \, + \, \beta_{1,j} s \, + \, \beta_{2,j} \phi \, + \, \beta_{3,j} \epsilon \, + \varepsilon \, ,
\label{eq:multilinear_fit}
\end{equation}
where $\nu_j$, $j=1,...,8$ represents the nestedness metrics indexed by $j$, $\beta_{0,j}$ is the intercept and $\beta_{i,j}$, $i = {1,..,3}$ are the partial regression coefficients. The $\varepsilon$ represents an error term. This sort of regression informs on the effect of a single network parameter when the rest of parameters are kept fixed. Such consideration is specially important given that, in natural systems, networks' properties are often correlated (for instance, larger networks tend to be less dense) and therefore bi-variate regressions may misleadingly quantify the influence of a certain property due to the uncontrolled coupled influence of another one. On the other hand, our model assumes a linear relation among the variables which might not always be accurate. Fig.~\ref{fig:correlations}b shows the results of the regression for each nestedness metrics, in particular, the significance of the partial coefficients corresponding to the different network parameters as well as the value of the adjusted coefficient of multiple determination (see the Methods section for more details).

\begin{figure}[h!]
\center
\includegraphics[width=0.6\textwidth]{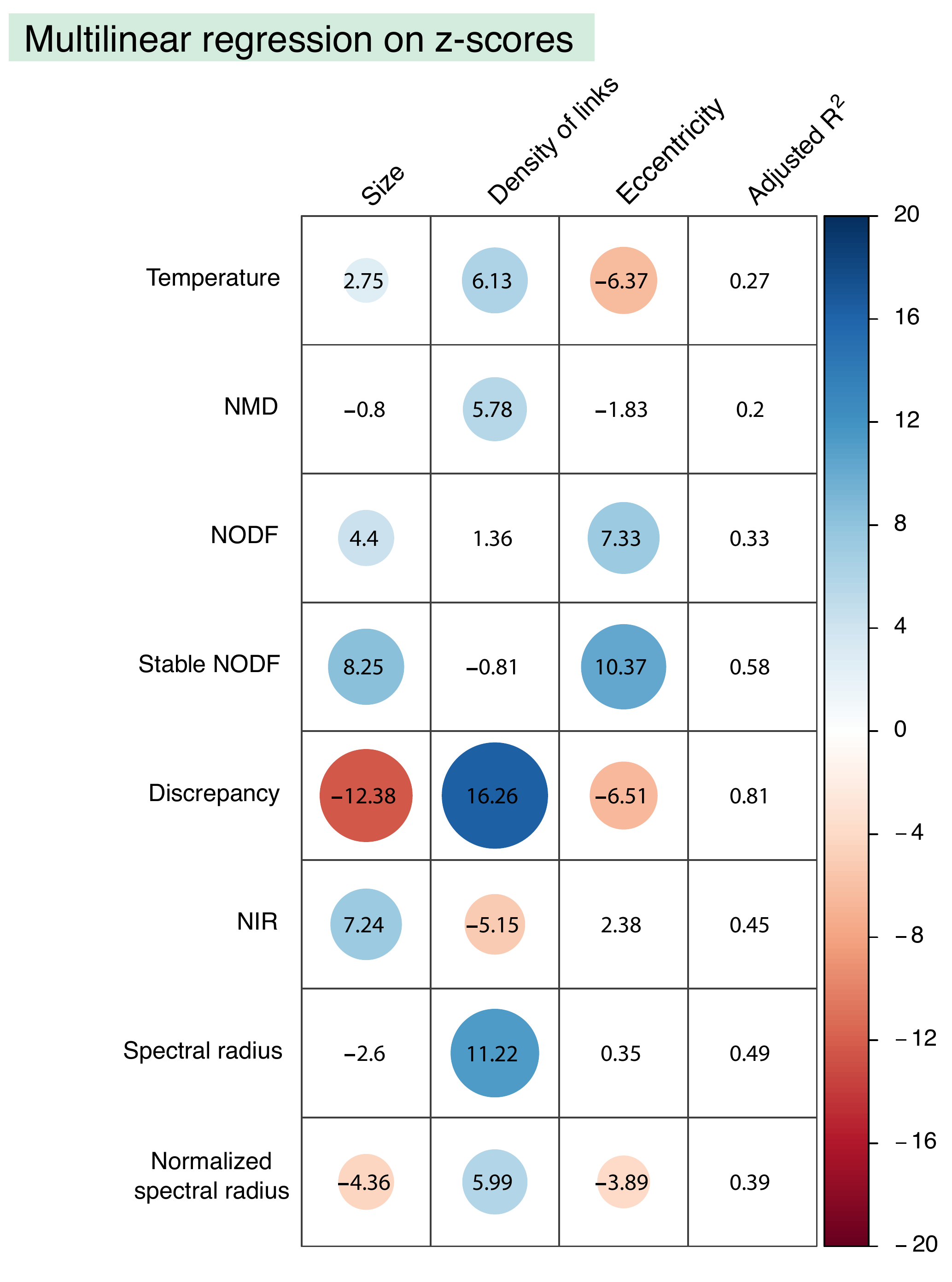}
\caption{\textbf{Dependency of z-scores on network parameters.} The figure summarizes the results of a multi-linear regression between the z-scores values corresponding to each nestednes metrics and network properties. Each row corresponds to the z-scores obtained by applying to each metrics the null model discussed in section~\ref{Null model}. The first column from the right shows the adjusted coefficient of multiple determination (\textit{adjusted} $R^2$). The other three columns show the \textit{t-ratio} of the regression coefficient corresponding to each explanatory variable (as labelled by the column name). Only those coefficients that are statistically significant ($p$-value $<0.01$) are highlighted by a colored circle, being the size and the color of the circle proportional to the $t$-ratio.}
\label{fig:correlations_zscore}
\end{figure}

Once we have quantified the dependencies of the various nestedness metrics on different network parameters, we next explore whether we can explain the deviations with respect to the null model observed in Fig.~\ref{fig:null_model}. In particular, we perform a multi-linear fit of the type detailed in Eq.~\ref{eq:multilinear_fit}, where we replace the nestedness values by the \textit{z-scores} obtained for each metrics when applying the null model discussed in sections~\ref{Null model} and~\ref{Results:null_model}. Such z-scores are calculated as follows:

\begin{equation}
\text{z-score}_j = \frac{\nu_j - \langle \nu_j \rangle }{\sigma_j} \, ,
\label{eq:zscore}
\end{equation}

where $\nu_j$ represents, as before, the real values obtained with a nestedness metrics indexed by $j$, $\langle \nu_j \rangle$ represents the average nestedness value calculated with metrics $j$ over the null ensemble and $\sigma_j$ represents the standard deviation of the distribution of nestedness in the ensemble for the same metrics. By fitting a linear function analogous to Eq.~\ref{eq:multilinear_fit} we obtained, thus, the partial coefficients which account for the contribution of each network parameter to the corresponding z-scores. A summary of these results can be found in Fig.~\ref{fig:correlations_zscore}.

\subsection{Critical analysis of each metrics}

In the previous section, we have quantified the influence of several network properties on various nestedness metrics, taking into account how each metrics measures the nestedness of empirical networks as well as how they compare to the null model of section~\ref{Null model}. With this information at hand, we now proceed to critically evaluate the performance of each metrics by discussing and framing the observed dependencies in a general context .

\subsubsection*{Temperature}

Despite its popularity, this metrics was already known to have several flaws~\cite{almeida2008consistent} and various authors have outlined the presence of ambiguous steps in its calculation~\cite{rodriguez2006new,mariani2019nestedness}. Indeed, Almeida-Neto et al.~\cite{almeida2008consistent} called up on its dependency on the density of contacts, $\phi$, and on the size of the matrix. We confirmed these dependencies since our statistical analysis shows that real values correlate positively with size and negatively with the density of links (see Fig.~\ref{fig:correlations}). Interestingly enough, the temperature is as well the only metrics to show a significant positive correlation with the degree degeneracy, while the rest of metrics penalize the repetition of degrees.

Moreover, when tested against the null model, the temperature exhibits a clear bias (see Fig.~\ref{fig:null_model}). In fact, the average nestedness in the ensemble is systematically larger than the real observations. The multilinear regression performed using the z-scores shows that they correlate significantly with the size, the density of links and the matrix eccentricity.  As shown in Fig.~\ref{fig:null_model}, the bias of this metrics shows a negative z-score value, therefore its modulus (which gives  the relative  distance to the  identity curve) increases when the size of the network is smaller, less dense and more eccentric. 
 Given that mutualistic ecological networks usually present low density and a pronounced  eccentricity, these conclusions point out that the temperature metrics should be applied, if at all used, with care in the ecological context.

\subsubsection*{NMD}

Our analysis of the nestedness metrics based on the Manhattan distance shows that it correlates positively with size and negatively with the degree degeneracy. Interestingly, the deviations with respect to the null model are sensibly smaller than in the temperature metrics, though a slight but systematic positive deviation still appears leading again to a negative z-score. The multilinear regression indicates that the z-scores are mainly explained by the density of links, which have a positive influence meaning that denser networks fall closer to their null expectation.

Overall, the NMD metrics exhibits notably less dependencies than its close metrics temperature, which with the NMD shares a common spirit given that both metrics measure somehow the distance of unexpected interactions. This dissimilarity is probably due to the different normalization of the NMD given in Eq.~\ref{eq:manhatan}. On the other hand, such normalization is dependent on the null model used~\cite{corso2008new}, and hence the metrics is inevitably subject to the same limitations (see Methods for more details on the implementation of NMD, in our case using the FF null model).

\subsubsection*{NODF and stable NODF}

In the work in which the NODF metrics was firstly proposed, Almeida-Neto et al.~\cite{almeida2008consistent} found a positive dependency with the matrix fill. In our analysis, we recover this result and observe as well a negative correlation with the network size (see Fig.~\ref{fig:correlations}a) which is nonetheless a veiled consequence of the variation in the density of links, as can be understood after performing the multilinear regression (see Fig.~\ref{fig:correlations}b). Furthermore, this nestedness index exhibits a good agreement with the null prediction, as was already found in~\cite{payrato2019breaking}. The differences with the null model, quantified by the z-score, are explained mainly by the size and the eccentricity. As expected from a statistical point of view, the small and eccentric networks show the largest difference with the null expectation. 

Although the NODF metrics is nowadays extensively used, some authors have raised a few concerns about its adequacy. In particular, Staniczenko et al.~\cite{staniczenko2013ghost} criticized the decreasing fill factor in its definition, which penalizes degree degeneracy. Indeed, we do observe a strong negative correlation with degree degeneracy for NODF in Fig.~\ref{fig:correlations}a. As a solution, Mariani et al.~\cite{mariani2019nestedness} proposed an alternative version of the metrics called stable-NODF, which does not incorporate this decreasing fill. Our analysis determines that dependencies of both versions of the metrics are very similar: on the one hand, the stable-NODF does moderate the correlations exhibited by NODF both on degree degeneracy and density of links, but on the other hand, the correlations of the z-scores with the size and eccentricity are strengthened.

\subsubsection*{Discrepancy}

The discrepancy index shows a significant dependency on the size and the density of links, being the latter parameter the dominant one as it can be seen from the multilinear regression (see Fig.~\ref{fig:correlations}). These dependencies had been noted already~\cite{almeida2008consistent}. Interestingly, these findings are very similar to the correlations observed for NODF and stable-NODF, despite the fact that the metrics are based on distinct strategies for measuring nestedness. 

On the other hand, the test against the null model reveals that for an important fraction of networks the real value of nestedness is smaller than the average in the ensemble, resulting in a systematic deviation a with negative z-score. This shift is very well explained by the regression of the z-scores summarized in Fig.~\ref{fig:correlations_zscore}, where it can be observed that the three network parameters studied correlate significantly with the z-scores. Indeed, the larger, less dense and more eccentric the network, the more distance there is between the null expectation of nestedness and the empirical value.

\subsubsection*{NIR}

The nestedness index based on network robustness exhibits no dependencies on the network parameters. Indeed, our statistical analysis reveals no significant correlation with any of the studied properties (see Fig.\ref{fig:correlations}). This suggests that, despite not being particularly popular, the NIR metrics is a reliable option for measuring nestedness. 
At the same time, the analysis done using the null model indicates that the nestedness value of smaller and denser networks tends to fall further apart from their null expectation. Indeed, this is a consequence of its definition, which relies on the difference between the areas of the ATCs obtained by the DDR and IDR node removal strategies. As the curvature of the former reproduces the shape of the IPN, it becomes less convex as the density increases, leading to a loss of sensitivity of this index. Therefore, this metrics is well adapted for ecological networks that usually show low densities, but less suited for other bipartite networks, like the aggregated market networks.

\subsubsection*{Spectral radius}

Among the metrics described, the spectral radius shows a significant dependency on both the size and the density of links, specially the former one. Indeed, larger and denser networks tend to have a larger spectral radius. This is a consequence of the lack of normalization, as mentioned in section~\ref{Metrics}. However, the spectral radius shows a remarkable agreement between the average over the ensemble and the value of the corresponding real network, along with a very low dispersion. Nonetheless, the z-scores correlate significantly well with the density of links, being the most denser networks the ones that exhibit a larger discrepancy with the null model.

In order to hinder the strong dependency on the network size, we evaluate as well a normalized version of the spectral radius. In particular, we weight the nestedness of each network with respect to the maximally ordered matrix with the same parameters as explained in the Methods section. Taking into account this normalization, it is now possible to compare the degree of nestedness of networks of different sizes, and to study how different network parameters affect the nestedness index. We find that this normalized version of the spectral radius correlates positively with the density of links, and negatively with the size and the eccentricity of the matrix. Notably, these dependencies are analogous to the ones shown by the NODF, stable NODF and discrepancy indexes. At the same time, the analysis against the null model reveals a slight deviation towards a larger value of the average in the random ensemble with respect to the empirical value. This deviation is stronger for larger, less dense and more eccentric networks.

Besides the mentioned dependencies, when using the spectral radius, it is essential to consider its underlying basis for measuring nestedness. As we pointed out in the introduction, the relation between the spectral radius and the degree of nestedness is not strictly monotonic, but only holds on statistical terms. This hampers its usefulness to rank networks according to their nestedness.

\section{Methods}
\label{Methods}

This section provides more details about the methodology used throughout our study. It contains as well information about the documented repository, called \emph{nullnest}, where we publish a set of codes to calculate nestedness using the studied metrics together with its null expectation. The present section is divided into the following parts:

\begin{itemize}
\item Dataset 
\item Generation of a sample of null networks
\item Computation of the nestedness index for each of the studied metrics
\item Statistical calculations
\item The \emph{nullnest} repository
\end{itemize}

\subsection{Dataset}
\label{Dataset}

Our study has been carried out using a large dataset composed by a total of 199 bipartite networks, including both ecological and economic interactions. The ecological networks were extracted from the Web of Life public repository~\cite{weboflife}, which contains the following data: 

\begin{itemize}
\item 118 Plant-pollinator networks. Here, the links among guilds represent mutualistic relationships, characterized by benefiting both interacting agents. In this case, animals, including mainly insects, pollinate flowering plants. This activity provides the pollinators with nutrients while pollinated plants enhance its reproductive success. The two set of nodes of the bipartite network represent the species of the plant and pollinator guilds.
\item 23 Seed-disperser networks. Here, the links also
represent, mutualistic interactions,  consisting now of birds feeding on the fruits of certain plants and then contributing to their reproduction and dispersal by disseminating their seeds. Hence, one guild is formed by the plant species and the other by the bird species.
\item 43 Host-parasite networks. Here, the links depict a parasitic relationship, where one of the species obtains benefits in detriment of the other. Explicitly, these networks are formed by different flea species which feed on diverse mammal species. Although this is not a mutualistic interaction, the system may still be represented by a bipartite network where the two guilds correspond to flea and mammals species.  
\item 4 Plant-herbivore networks. Here, the links represent a consumer-resource interaction between insect species (one guild) and plant species (the other guild). In detail, the networks depict different communities where macrolepidopteran species feed on several Prunus species. 
\item 3 Plant-ant networks. These networks include two examples of diverse types of communities: a network depicting ants which feed on plant nectar, this being a consumer-resource interaction, and two networks representing communities where ant species live in a mutualistic association with certain plant species known as Myrmecophytes.  
\end{itemize}

On the other hand, our dataset includes a number of economic networks which are publicly available in~\cite{hernandez2018figshare}, consisting of:

\begin{itemize}
\item 8 economic networks representing buyers-sellers interactions in the Boulogne-sur-Mer Fish Market in France~\cite{hernandez2018trust}. These are  mutualistic networks taken from  a very different context. Each network describes the transactions observed in different days in the bilateral or in the auction Fish Market. These daily networks are typically much denser than ecological ones.
\end{itemize} 

All the networks in our dataset where treated as binary (non-weighted links). We have only kept in our study networks with a minimum size of 20 nodes.

\subsection{Generation of a sample of null networks}
\label{Methods_null}

In order to maximize the entropy under the constraint of keeping the average degree sequence fixed and equal to the observed one, it is necessary to apply the  Lagrange Multipliers' technique. The determination of the value of such multipliers, obtained by maximizing the likelihood that the empirical degree sequence appears with maximum probability in the random ensemble, provides, as aforementioned, the probability of interaction among species or agents from different guilds. This probabilistic bipartite matrix can then be used to sample the random ensemble of networks.

For each real network in our dataset, we sampled $10^4$ null networks with the obtained probability interaction matrix. Across the same sample, each of these null matrices may vary in its size (number of connected nodes), density of links, degree sequence, redundancy of degrees or bipartite matrix excentricity. Nevertheless, the average degree sequences are maintained equal to the empirical ones.    

\subsection{Computation of the nestedness index for each of the studied metrics}
\label{Methods_metrics}
\subsubsection*{Temperature}

We calculated the temperature metrics using the \textit{R} software~\cite{RCore} and, specifically, the \textit{bipartite} package~\cite{dormann2009indices,BipartitePackage1} version 1.13.0. In particular, we used the \textit{nested} function and we set as method the \textit{binmatnest2} option. This calculates the temperature metrics by using an implementation by Jari Oksanen~\cite{veganpackage} of the \textit{binmatnest} program by Miguel Rodriguez-Girones~\cite{rodriguez2006new}. 
We have redefined the resulting temperature, as shown in eq.~(\ref{eq:redef_T}) in order to uniform the interpretation of the values yielded by all the metrics such that the higher the value of the corresponding index, the higher the nestedness.

\subsubsection*{NMD}
 
We calculated the nestedness metrics based in the Manhattan distance (NMD) using the \textit{R} software~\cite{RCore}. We used the \textit{nestedness.corso} function (currently deprecated) from the \textit{bipartite} package~\cite{dormann2009indices,BipartitePackage1} version 0.90. For each measure (both for the real networks and the sampled networks), we set the number of null networks that eventually permits evaluating the significance to 500.  Again, we redefined the index as shown in Eq.~\ref{eq:redef_NMD} to simplify the interpretation of results.

\subsubsection*{NODF}

We wrote a program in FORTRAN90 that computes the NODF and stable-NODF metrics for the real networks, as well as for the corresponding set of null networks.

Importantly, when performing the calculations over the sample of null networks, we kept the same normalization for all sampled networks. That is, we divided the number of overlapping connections, calculated for each null network, by the original number of rows and columns, independently of whether some of the nodes came to have zero degree in the null network.  

\subsubsection*{Discrepancy}

We computed the discrepancy metrics using the \textit{R} software~\cite{RCore} and the \textit{bipartite} package~\cite{dormann2009indices,BipartitePackage1} version 1.13.0. In particular, we performed the calculation using the method \textit{discrepancy} from the \textit{nested} function. 

The final nestedness value measured is directly proportional to the density of links and size of the network. With the aim of removing such dependencies, we divided the resulting value of the metrics by the total numbers of links. This results in a relative discrepancy. Once again, we finally applied Eq.~(\ref{eq:redef_dis}) to obtain the same monotonic variation for all the indices.

\subsubsection*{NIR}

We implemented a program in FORTRAN90 that calculates the NIR value of the real network and of the corresponding set of null networks. In each case, the resulting value of nestedness is multiplied by 100 in order to preserve the same scale for all the metrics.

In order to account for the possible effects of the degeneracy in the ordering, that is, the fact that multiple configurations are possible when we order rows and columns by their degree, we computed the resulting NIR as the average over a large number of equivalently ordered configurations. These configurations were produced by randomly swapping the matrix position of nodes with the same degree. In more detail, to generate a new ordering we run over all the nodes with degenerate degree and, for each node, we accept a position swap with probability $\frac{1}{2}$. 

For each real network, we calculated the degeneracy, \textit{ideg}, the number of repeated degrees. Then, we produced a total of $10 \cdot ideg$ configurations with the same degree order but diverse row and column positions. This procedure was carried out both for the real network and for each null network in the sampling with the exception of the Robertson's network~\cite{robertson1929flowers} for which, due to its very large size (1500 species), only 10 degenerate configurations have been computed.

\subsubsection*{Spectral radius}

We computed the largest eigenvalue using the \textit{R} software~\cite{RCore}, in particular the \textit{eigs\_sym} function from the \textit{rARPACK} package~\cite{rarpack}.

In order to calculate the normalized version of this metrics, we need an estimation of the largest spectral radius of a perfectly nested network of the  same size and fill. To estimate each of these values, for each real network in our dataset we produced 100 new networks, characterized by being perfectly nested. These networks were generated using the SNM algorithm~\cite{burgos2007nestedness}, which preserves the number of connected nodes (network size) and links, but modifies the pattern of connections and the degree sequences. This algorithm is divided into two procedures. First, the real network is randomized preserving only the fill and the size (that this, ensuring that every node has at least one connection). Second, the SNM algorithm is performed, which consist of iterating the following rules: 

\begin{enumerate}[(i)]
\item We attempt to modify a link by proposing a new partner, randomly selected but different to the original node. The rewiring is susceptible of being accepted only if the new partner has a larger degree than the previous one. This step performs a static version of preferential-attachment. 
\item If the proposed reconnection leaves one of the nodes with zero degree, the move is discarded. This ensures that the number of connected nodes does not change, thus preserving the network size.
\end{enumerate}

By iterating over these steps \textit{i} and \textit{ii}, one generates a new matrix which is more nested as well as more heterogeneous in its degree sequences than the original one (see~\cite{payrato2019breaking}). The iteration stops when no more moves are allowed. However, given condition \textit{ii}, this process is not unique and might end up in multiple perfectly nested configurations. To handle this, we generated several optimal configurations per each real network. Specifically, we generated 100 new networks for each empirical network, and exceptionally, for computational reasons, 50 networks for the very large Robertson network. This means that Eq.~\ref{eq:spectral_radius} is actually calculated as:
\begin{equation}
\rho_{norm} \, = \, 100 \, \, \frac{\rho}{\sum _i ^{100} \frac{\rho _{perfect,i}}{100} }, 
\label{eq:spectral_radius_average}
\end{equation}

where $\rho$ is the spectral radius of the real network and $\rho _{perfect,i}$ represents an optimal configuration with the same size and fill of the real network, produced by the SNM algorithm.

When sampling the ensemble, we generated 10 nested networks per each null network, and in order to keep the calculations computationally feasible we reduced the sampling size to 500 null networks. Accordingly, the average normalized spectral radius is calculated as:
\begin{equation}
\langle \rho_{norm} \rangle \, = 100 \, \sum_j^{500} \, \, \frac{\rho_{null,j}}{500 \, \sum _i ^{10} \frac{\rho _{perfect,i,j}}{10}} \, ,
\label{eq:spectral_radius_null}
\end{equation}

where $\rho_{null,j}$ represents a null network sampled from the statistical ensemble and $\rho _{perfect,i,j}$ represents a perfect configuration produced with the SNM algorithm, having the same size and fill as the corresponding null network.

\subsection{Statistical calculations}

The statistical correlations were numerically calculated using Python. The Spearman rank correlation coefficient $r_s$ and its $p$-value were calculated using the Scipy package~\cite{scipy}, in particular the \textit{scipy.stats.spearmanr} function. 

We performed the linear fits using the \textit{Statsmodels} package~\cite{seabold2010statsmodels}, which carries out a multi-linear least-square regression and provides multiple information, including the adjusted $R^2$, the partial regression coefficients, their standard deviation and their associated $p$-value. The $t$-ratio$_{i,j}$ corresponding to each partial regression coefficient, $\beta_{i,j}$, is calculated as follows:
\begin{equation}
t-\text{ratio}_{i,j} = \frac{\beta_{i,j}}{\sigma_{i,j}}
\label{eq:tratio}
\end{equation} 

where $\sigma_{i,j}$ is the standard deviation associated to that coefficient. This index provides, hence, information on how significantly different from zero is a certain regression coefficient.
 
\subsection{The \emph{nullnest} repository}

Additionally to the comparative study here presented, we provide an open github repository named \emph{nullnest} that aims at being a practical tool, for both ecologists and network scientist, to assess the nestedness of real and null networks. The repository is thoroughly documented, with examples and ready-to-use programs, and allows performing the analysis discussed in the present paper as well as some of the main calculations derived in~\cite{payrato2019breaking}.  

The \emph{nullnest} repository is divided into two main blocks. The first part is related to the construction of the null model. In particular, we give a program to compute the null model discussed in section \ref{Null model} for any bipartite network introduced by the user. We also provide the ready-to-use probabilities of interaction in the null ensemble, for the whole dataset of empirical networks described in section~\ref{Dataset}. On the other hand, the second part of the package is concerned with the measurement of nestedness. In detail, we provide the codes to quantify the real degree of nestedness of a given matrix, together with the first two moments of its null distribution, using any of the six metrics discussed here. This can be done either by using the analytical expressions derived in~\cite{payrato2019breaking} (only for NODF and the spectral radius) or by numerically sampling the null ensemble as explained in subsections~\ref{Methods_metrics} (for any of the six metrics).

\section{Discussion and Conclusions}
\label{Conclusions}

Although it has recently been shown that nestedness is not an emergent irreducible property of the network,  it still remains an interesting quantity to measure, as it constitutes a global property that informs on the heterogeneity of the degree distributions of the guilds. This is particularly relevant for ecological networks because of their typical, rather small sizes preclude a correct fit to a fat tail distribution, like a power law,  on the available data. Because of the interest in ecology for this property,  different definitions of nestedness coexist in the literature. These metrics usually quantify some property of the network following a precise protocol, leading to  {\em operational definitions}. Moreover, several of these metrics are integrated into packages widely used by network ecologists to assess the nestedness values of different networks. 
This lack of a unique definition generates confusion when it comes to the comparison between the nestedness values of different networks. 

In this work, we have performed a systematic comparative study of the performances of six different metrics and the variants of two of them, addressing their dependency on various network parameters. Based on a large database of real systems, our results clearly put in evidence that the different metrics show diverse dependencies on size, density of contacts, eccentricity and degree degeneracy. Therefore, if the same group of networks is ranked according to their nestedness, the outcome will depend on the metrics used. Understanding these dependencies for each metrics has helped us to explain, as well, the systematic shifts between the real values of nestedness and the average over a null model based on a maximum-entropy, maximum-likelihood ensemble.

The nestedness metrics studied here may be roughly classified in three groups according to the properties of the networks that are used to define them: (i) nestedness metrics based on the number of misplaced elements in the bipartite adjacency matrix with respect to a perfectly nested matrix, like Discrepancy,  (ii) nestedness metrics based on global properties of the network like, NODF, NIR, and $\rho$ and (iii) nestedness metrics like T and NMD that operate similarly to (i) but weighting the distance of misplaced interactions to their ideal location in the perfect nested matrix. Our results point out that the NIR index is, by far, the most independent metrics with respect to the considered network parameters,  although it suffers from a lack of sensitivity when the density of contacts is high. Moreover, the NODF, the stable NODF, the discrepancy and the normalized spectral radius all show very similar dependencies, that is: a positive correlation with the density of links and, for the latter two, a negative correlation with the size. While a dependency with the size is undesired and ought not to appear when using a proper normalization, some authors have claimed that a positive correlation between nestedness and fill is in fact expected~\cite{almeida2008consistent}.

Our work aims at providing a useful guide addressed at practitioners that compiles the different characteristics, advantages and drawbacks of the most popular nestedness metrics. We also extended the use of maximum-entropy-based null models~\cite{payrato2019breaking,squartini2011analytical,saracco2015randomizing} to these metrics. Finally, this work is accompanied by a package that allows to calculate all the nestedness indicators studied, generate the null ensemble for any network, as well as a database with the already calculated probabilities allowing to generate the null models for the 199 networks studied here. 

\section*{Acknowledgements}

Y. M. acknowledges partial support from the Government of Aragon, Spain through grant E36-17R (FENOL), by MINECO and FEDER funds (FIS2017-87519-P) and from Intesa Sanpaolo Innovation Center. C. P. B. acknowledges support of the LABEXMME-DII (Grant No. ANR reference 11-LABX-0023). The funders had no role in study design, data collection, and analysis, decision to publish, or preparation of the manuscript. 

\section*{Data availability}

The codes used for the analysis presented throughout this paper, as well as the main results of the null model concerning the real networks studied, will be made public as a \textit{github} repository under the name \textit{nullnest}. The real networks analyzed are already public in the Web of Life site~\cite{weboflife} and a figshare repository~\cite{hernandez2018figshare}. 


\end{document}